\documentclass[journal]{IEEEtran}

\ifCLASSINFOpdf
\else
   \usepackage[dvips]{graphicx}
\fi
\usepackage{url}

\usepackage{graphicx}
\usepackage{amsmath}
\usepackage{amsfonts}
\usepackage{bm}
\usepackage{subfigure}

\begin{document}

\title{Enhancing and Learning Denoiser without Clean Reference}

\author{Rui Zhao, Daniel P.K. Lun, and Kin-Man Lam
\thanks{Department of Electronic and Information Engineering, The Hong Kong Polytechnic University, Hong Kong.}
\thanks{rick10.zhao@connect.polyu.hk}
}

\markboth{Journal of \LaTeX\ Class Files, Vol. **, No. *, Month Year}
{R. Zhao \MakeLowercase{\textit{et al.}}: NTGAN: Learning Blind Image Denoising without Clean Reference}
\maketitle

\begin{abstract}
Recent studies on learning-based image denoising have achieved promising performance on various noise reduction tasks. Most of these deep denoisers are trained either under the supervision of clean references, or unsupervised on synthetic noise. The assumption with the synthetic noise leads to poor generalization when facing real photographs. To address this issue, we propose a novel deep image-denoising method by regarding the noise reduction task as a special case of the noise transference task. Learning noise transference enables the network to acquire the denoising ability by observing the corrupted samples. The results on real-world denoising benchmarks demonstrate that our proposed method achieves promising performance on removing realistic noises, making it a potential solution to practical noise reduction problems.
\end{abstract}

\begin{IEEEkeywords}
Learnable gradient-based cutout, data augmentation, deep neural network regularization.
\end{IEEEkeywords}

\IEEEpeerreviewmaketitle

\section{Introduction}
\label{sec:intro}
Convolutional neural networks (CNNs) have been widely studied over the past decade in various computer vision applications, including image classification \cite{Krizhevsky2012ImageNetCW}, object detection \cite{Ren2015FasterRT}, and image restoration \cite{Zhang2017BeyondAG}. In spite of high effectiveness, CNN models always require a large-scale dataset with reliable reference for training. However, in terms of low-level vision tasks, such as image denoising and image super-resolution, the reliable reference data is generally hard to access, which makes these tasks more challenging and difficult. Current CNN-based denoising methods usually approximate the realistic noise as the additive white Gaussian noise (AWGN), and synthesize the pairs of the corrupted observation $\mathbf{y}_i$ and the clean reference $\mathbf{x}_i$ for training. With a large amount of these synthetic pairs, a deep regression model, i.e. a convolutional neural network, can be trained by minimizing the empirical risk, as follows:
\begin{equation}
    \Theta^{*} = \arg\min_{\Theta}\sum_{i}\mathcal{L}(f(\mathbf{y}_i; \Theta), \mathbf{x}_i),
    \label{eq:1}
\end{equation}
where $f$ denotes a mapping model from $\mathbf{y}_i$ to $\mathbf{x}_i$ with its trainable parameters $\Theta$ under a loss function $\mathcal{L}$. However, in some practical applications, e.g., medical imaging and hyperspectral remote sensing, the acquisition of noise-free data is expensive, or even impractical.

In addition, most state-of-the-art image denoising methods require \textit{a priori} knowledge of the noise level when performing inference. Methods, based on image prior modeling \cite{Gu2014WeightedNN,Li2018PredictiveCM}, usually involve some optimization processes, and the hyperparameters for solving these optimization problems are highly related to the noise level. On the other hand, methods, based on plain discriminative learning, attempt to either learn a denoising function for a specific noise level, e.g., DnCNN \cite{Zhang2017BeyondAG}, or take the noise level as an input variable when training the models, e.g., FFDNet \cite{Zhang2017FFDNetTA} and CBDNet \cite{Guo2018TowardCB}. Therefore, the noise level plays an important role in practical noise reduction tasks. However, current deep unsupervised image denoisers lack the utilization of the information provided by the noise level, which results in poor generalization and flexibility when solving practical denoising problems.

In this paper, we propose a novel noise transference generative adversarial network, namely NTGAN, to tackle the problems of limited reference data and poor generalization in denoising real-world photographs. Specifically, NTGAN attempts to learn the noise transference between the paired noisy observations. We follow the idea of conditional generative adversarial network \cite{Mirza2014ConditionalGA} and establish a controllable generator for different noise levels.

In summary, the main contributions of this work can be concluded as follows:
\begin{itemize}
    \item We treat noise reduction as a special case of noise transference, and establish a conditional regression model with the guidance from the target noise level, which enables the proposed denoiser to be trained by observing corrupted samples only.
    \item We present a strategy for generating paired training samples from real photographs with the pseudo supervision from the target noise level maps, which allows the network to obtain better generalization ability for removing realistic noises.
\end{itemize}

\section{Related work}
\subsection{Blind image denoising}
Blind image denoising is a challenging and complicated task, because the realistic noise is spatially variant and signal-dependent, which makes the noise distribution hard to estimate. Methods, based on image prior modeling, usually approximate the realistic noise as AWGN and estimate the noise level via principal component analysis \cite{Chen2015AnES,Liu2013SingleImageNL,Pyatykh2013ImageNL} and wavelet transform \cite{Portilla2004FullBD} for effective noise removal. Recently, more and more attention has been paid to the extension of AWGN. Numerous studies have been proposed to investigate the noise correlations between image channels \cite{Xu2017MultichannelWN} and image patches \cite{Xu2018ATW}. On the other hand, methods, based on plain discriminative learning, are generally trained with the synthetic pairs based on AWGN \cite{Chen2017TrainableNR,Zhang2017BeyondAG}. A great breakthrough comes from a two-stage denoising system, i.e. CBDNet \cite{Guo2018TowardCB}, which aims to estimate the pixel-wise noise level via a sub-network, followed by a non-blind deep denoiser for removing the noise. VDNet \cite{NIPS2019_8446} further enhances the noise model in CBDNet, and achieves the state-of-the-art performance. RIDNet \cite{Anwar_2019_ICCV}, from another aspect, proposed that the ``residual-on-residual'' architecture can perform blind image denoising in a one-stage manner.
\subsection{Unsupervised image denoising}
In spite of their great power of removing noise, the learning-based methods usually require a large amount of noisy and noise-free image pairs for training. However, the noise-free references are generally difficult to obtain in practice, which leads to the challenge of unsupervised image denoising. Noise2Noise (N2N) \cite{Lehtinen2018Noise2NoiseLI} proposed that the models, learnt from the paired corrupted observations, can be applied to effectively removing the noise. In addition, Noise2Self (N2S) \cite{N2S} and Noise2Void (N2V) \cite{N2V} adopted the self-supervision strategy to train a regression model for blind noise removal. From another point of view, the work of deep image prior (DIP) \cite{Ulyanov2017DeepIP} showed that the network architecture itself contains \textit{a priori} knowledge, which can be used for image restoration. However, the above-mentioned works mainly consider the synthetic noise, which results in a dramatic degradation of their performance when facing real photographs. Therefore, in this paper, we move one step further to establishing an unsupervised denoising framework for the realistic noise.

\section{The proposed noise transference GAN}
In this section, we present the proposed NTGAN for blind unsupervised image denoising. Firstly, we analyze the relationship between noise transference and noise reduction. Secondly, we introduce the proposed pseudo noise level map for unsupervised learning. Finally, we present the learning strategy for training the proposed NTGAN.
\subsection{Noise transference} \label{sec:3.1}

Conventional deep denoisers \cite{Zhang2017BeyondAG,WDnCNN} attempt to learn a denoising function for a specific noise level. Therefore, the resultant model parameters $\Theta$ become a function of the noise level $\sigma$ as $\hat{\mathbf{x}} = f(\mathbf{y}; \Theta(\sigma))$, where $\hat{\mathbf{x}}$ is the estimated noise-free image, and $\mathbf{y}$ is the noisy observation with the noise level $\sigma$. As claimed in FFDNet \cite{Zhang2017FFDNetTA}, the model flexibility can be enhanced by introducing the noise level $\sigma$ as an input variable into the model. Thus, the denoising function becomes $\hat{\mathbf{x}} = f(\mathbf{y}, \sigma; \Theta)$. During training, the following objective function is minimized:
\begin{equation}
    \Theta^{*} = \arg\min_{\Theta}\sum_{i}\mathcal{L}(f(\mathbf{y}_i, \sigma_i; \Theta), \mathbf{x}_i),
    \label{eq:4}
\end{equation}
where $\mathbf{x}_i$ represents the target clean reference. Thus, the resultant $\Theta^{*}$ becomes independent of the noise level $\sigma$, because $f_{\Theta^{*}}$ is able to minimize the loss $\mathcal{L}$ in Eqn. (\ref{eq:4}) for arbitrary $\sigma$. However, $\Theta^{*}$ still depends on the noise level of the target image $\mathbf{x}_i$, which indicates that Eqn. (\ref{eq:4}) is equivalent to
\begin{equation}
    \Theta^{*} = \arg\min_{\Theta}\sum_{i}\mathcal{L}(f(\mathbf{y}_i, \sigma_{\mathbf{y}_i}; \Theta(\sigma_{\mathbf{x}_i})), \mathbf{x}_i)|_{\sigma_{\mathbf{x}_i} = 0},
    \label{eq:5}
\end{equation}
where $\sigma_{\mathbf{y}_i}$ and $\sigma_{\mathbf{x}_i}$ represent the noise level of the source and the target images, respectively. If we follow the transformation in FFDNet \cite{Zhang2017FFDNetTA} and take $\sigma_{\mathbf{x}_i}$ as an input variable of the mapping function, we can obtain a general form of Eqn. (\ref{eq:5}), as follows:
\begin{equation}
     \Theta^{*} = \arg\min_{\Theta}\sum_{i}\mathcal{L}(f(\mathbf{y}_i, \sigma_{\mathbf{y}_i}, \sigma_{\mathbf{z}_i}; \Theta), \mathbf{z}_i),
     \label{eq:6}
\end{equation}
where the target image $\mathbf{z}_i$ can be a corrupted image with the noise level $\sigma_{\mathbf{z}_i}$. Eqn. (\ref{eq:6}) implies that the resultant model becomes a noise transference network, which can be trained without any clean reference.

In addition, we mainly consider the realistic noise in this work, and thus the noise level of the input observation $\sigma_{\mathbf{y}_i}$ is not available for training. Therefore, we adopt the ``residual-on-residual'' \cite{Anwar_2019_ICCV} architecture to establish NTGAN, and omit the source noise level $\sigma_{\mathbf{y}_i}$ in Eqn. (\ref{eq:6}), as follows:
\begin{equation}
     \Theta^{*} = \arg\min_{\Theta}\sum_{i}\mathcal{L}(f(\mathbf{y}_i, \sigma_{\mathbf{z}_i}; \Theta), \mathbf{z}_i).
     \label{eq:7}
\end{equation}
Eqn. (\ref{eq:7}) indicates that NTGAN is a blind denoiser, which does not require \textit{a priori} knowledge on $\sigma_{\mathbf{y}_i}$ for noise removal. It is worth noting that the newly introduced input variable $\sigma_{\mathbf{z}_{i}}$ controls the noise level we expect NTGAN to synthesize, which is manually set during the noise generation process, and thus it is available for training. Moreover, during the denoising inference, the target noise level is set to zero as follows:
\begin{equation}
    \hat{\mathbf{x}} = f(\mathbf{y}, \sigma_{\mathbf{z}}; \Theta^{*})|_{\sigma_{\mathbf{z}} = 0},
    \label{eq:8}
\end{equation}
because we expect NTGAN to generate a noise-free ($\sigma_{\mathbf{z}} = 0$) image. On the other hand, the noise level of a real-world noisy image cannot be simply described by a single value $\sigma_{\mathbf{z}}$. Therefore, we follow the settings in CBDNet \cite{Guo2018TowardCB}, and establish the noise level map $\mathbf{M}_{\mathbf{z}}$ to replace $\sigma_{\mathbf{z}}$ in Eqns. (\ref{eq:7}) and (\ref{eq:8}), which consists of the pixel-wise noise standard deviations in the target observation $\mathbf{z}$.

In summary, NTGAN is a conditional generator for different noise levels. It takes a noisy observation $\mathbf{y}$ and a target noise level map $\mathbf{M}_{\mathbf{z}}$ as inputs, and produces a new observation $\mathbf{z}$ with the guiding (expected) noise level $\mathbf{M}_{\mathbf{z}}$.

\subsection{Pseudo target noise level map}
As introduced in Sec. \ref{sec:3.1}, NTGAN requires the noise level maps of the target noisy images for training. However, such a paired set, $[\mathbf{y}, (\mathbf{z},\mathbf{M}_{\mathbf{z}})]$, is hard to obtain when only noisy samples are available. As the noise level of $\mathbf{y}$ is not available, leading to the difficulty in obtaining true $\mathbf{M}_{\mathbf{z}}$. Therefore, we further propose a strategy for generating pseudo target noise level maps when training with the real-world observations. As the noise residing in real photographs is relatively weak compared to the image content, we can synthesize the pairs of observations by adding the synthetic noise to a real photograph $\mathbf{y}$ as follows:
\begin{equation}
    \mathbf{z} = \mathbf{y} + \mathbf{n},
    \label{eq:9}
\end{equation}
where $\mathbf{n}$ is the synthetic noise from a noise generation model. Thus, the target noise level map $\mathbf{M}_{\mathbf{z}}$ can be approximated as the noise level of $\mathbf{n}$, which can be obtained from the noise generation process. NTGAN aims to learn the transference between the paired observations with different noise intensities. Therefore, the noise residing in $\mathbf{y}$ and $\mathbf{z}$ should follow the same distribution. We assume that the realistic noise follows a heterogeneous Gaussian distribution, and so adopt the noise generation model from \cite{LiuNoiseEF2006}, which considers both the signal-independent and signal-dependent components in $\mathbf{n}$. In addition, we also apply the in-camera pipeline in \cite{Guo2018TowardCB} to the noise generation model to further narrow the domain gap between the noise in $\mathbf{y}$ and $\mathbf{z}$. Thus, the resultant noise generation model is formulated as follows:
\begin{equation}
    \begin{aligned}
        \mathbf{n}(\mathbf{y}, \mathbf{M}_{\mathbf{z}}) &= f_{\text{BPD}}(f_{\text{crf}}(\mathbf{L} + \mathbf{n}_{s}(\mathbf{L}) + \mathbf{n}_{c}))
        - f_{\text{BPD}}(f_{\text{crf}}(\mathbf{L})), \\
        \text{with }\mathbf{L} &= f_{\text{icrf}}(\mathbf{y}),
        \label{eq:10}
    \end{aligned}
\end{equation}
where $f_{\text{crf}}$ and $f_{\text{icrf}}$ represent the camera response function and the inverse camera response function, respectively. Specifically, $f_{\text{icrf}}$ transforms the original image $\mathbf{y}$ to obtain an irradiance plane $\mathbf{L}$. Moreover, $\mathbf{n}_{s}$ and $\mathbf{n}_{c}$ account for the noise components that are dependent and independent of the signal $\mathbf{y}$, respectively. Therefore, the pseudo target noise level map $\mathbf{M}_{\mathbf{z}}$ is defined as follows:
\begin{equation}
    \mathbf{M}_{\mathbf{z}}^2 = \mathbf{L}\cdot\sigma_{s}^{2} + \sigma_{c}^2,
    \label{eq:11}
\end{equation}
where $\mathbf{L}\cdot\sigma_{s}^{2}$ and $\sigma_{c}^{2}$ are the noise variance of $\mathbf{n}_{s}$ and $\mathbf{n}_{c}$, respectively. $f_{\text{BPD}}$ in Eqn. (\ref{eq:10}) represents the function considering the Bayer patterning and demosaicing, which was proposed in \cite{laroche1994apparatus} for simulating the spatially correlated noise. This noise generation model guarantees the statistics consistency of the noise in $\mathbf{y}$ and $\mathbf{z}$, because the resultant noise in both $\mathbf{y}$ and $\mathbf{z}$ is heterogeneous Gaussian distributed with the dependency on the clean signal. With the proposed strategy for the pseudo noise level map, the training pairs can be obtained from the noisy observations by using Eqns. (\ref{eq:9}), (\ref{eq:10}), and (\ref{eq:11}) to form the training set as $\{[\mathbf{y}_{1}, (\mathbf{M}_{\mathbf{z}_{1}}, \mathbf{z}_{1})]$ $, [\mathbf{y}_{2}, (\mathbf{M}_{\mathbf{z}_{2}}, \mathbf{z}_{2})], \dots, [\mathbf{y}_{n}, (\mathbf{M}_{\mathbf{z}_{n}}, \mathbf{z}_{n})]\}$. 
\subsection{Framework}
\begin{figure*}[t]
    \centering
    \includegraphics[width = 0.9\linewidth]{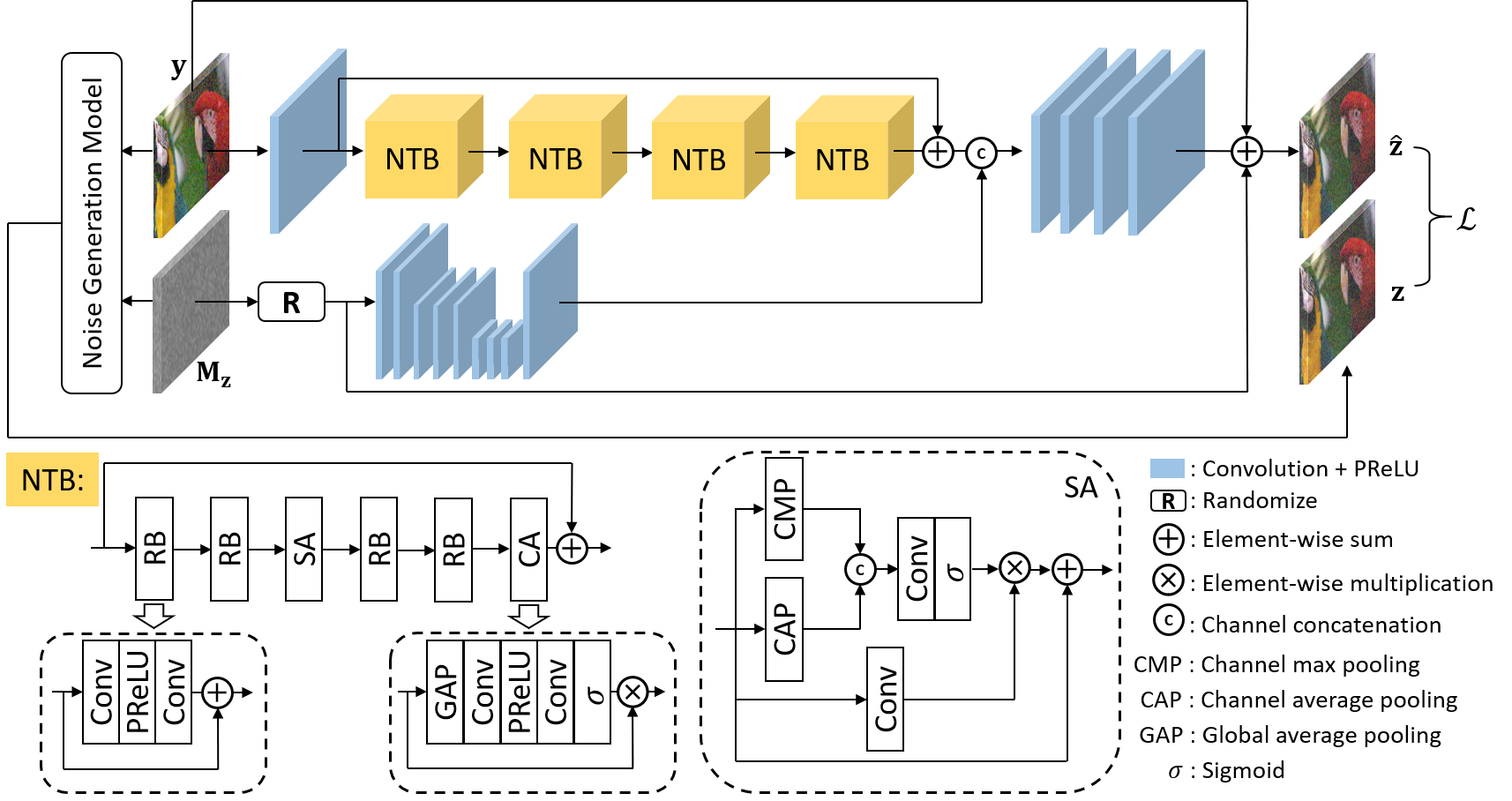}
    \caption{The proposed NTGAN for blind unsupervised image denoising. The last convolutional layers in both branches are not followed by PReLU. NTB: noise transference block, RB: residual block. SA: spatial attention unit. CA: channel attention unit.}
    \label{fig:1}
\end{figure*}

As introduced in Sec. \ref{sec:3.1}, we aim to introduce the noise level of the target image as an input variable of the model, so that the resultant network can be a conditional noise generator, controlled by $\mathbf{M}_{\mathbf{z}}$. In addition, the ``residual-on-residual'' structure is required to omit the supervision from $\sigma_{\mathbf{y}}$. Therefore, we establish the network as shown in Figure \ref{fig:1}.
Specifically, NTGAN consists of an image-encoding stream and a noise-level-encoding stream, which are presented as the top and the bottom branches in Figure \ref{fig:1}, respectively. The image-encoding branch consists of four noise transference blocks (NTB), each of which contains a spatial attention (SA) unit, a channel attention (CA) unit and four residual blocks (RB). The two attention units follow the designs in \cite{Anwar_2019_ICCV} and \cite{zhang2019rnan}, in order to focus on the important regions in both the spatial and channel dimensions. Specifically, in the SA unit, an input feature map $\bm{F} \in \mathbb{R}^{c\times h\times w}$ is first independently compressed by average pooling and max pooling in the channel dimension. The two compact feature maps are concatenated, and fed to a convolution layer with the sigmoid function to form a heat-map $\bm{H} \in \mathbb{R}^{1\times h\times w}$. We use $\bm{H}$ to rescale $\bm{F}$ along the spatial dimension for spatial attention. Similarly in CA, the global average pooling is first applied to $\bm{F}$ to reduce the feature size to $ \bm{v} \in \mathbb{R}^{c\times 1 \times 1}$. We generate the channel heat-vector by passing $\bm{v}$ to a non-linear mapping module with the sigmoid function. The produced heat-vector is used to rescale the different channels in $\bm{F}$ to achieve channel attention. All RB, NTB, and the global streams employ the residual connections, which inherits the `residual-on-residual'' structure in \cite{Anwar_2019_ICCV} to achieve noise transference without knowing the source noise level. The noise-level-encoding branch consists of convolutional layers, average pooling layers, and a randomization block. The randomization is defined as the element-wise multiplication with a set of standard Gaussian distributed random variable $\mathbf{r}\sim\mathcal{N}(0, 1)$, which follows the design in \cite{pmlr-v48-larsen16}. The features from the two streams are fused by the channel concatenation. Finally, we reconstruct the target observation $\hat{\mathbf{z}}$ ,based on the fused features, to mislead a patch-based discriminator.

To learn the noise transference task, we consider two loss terms in the objective function for training. NTGAN aims to produce a noisy observation with the expected noise level. Thus, the produced noisy image aims to fool the discriminator, while maintaining the image content. Therefore, the training objective is formulated as follows:
\begin{equation}
    \mathcal{L} = \mathcal{L}_{\text{GAN}} + \lambda \mathcal{L}_{\text{rec}}, 
    \label{eq:12}
\end{equation}
where $\mathcal{L}_{\text{GAN}}$ and $\mathcal{L}_{\text{rec}}$ are the loss of the discriminator and the reconstruction, respectively, and $\lambda$ is a hyperparameter controlling the trade-off between these two loss terms. Specifically, $\mathcal{L}_{\text{rec}}$ is computed as follows:
\begin{equation}
    \mathcal{L}_{\text{rec}} = \frac{1}{n}\sum_{i=1}^{n}||\hat{\mathbf{z}}_i - \mathbf{z}_i||^{2}_{2},
    \label{eq:13}
\end{equation}
where $\hat{\mathbf{z}}$ denotes the reconstructed observation from NTGAN, and $\mathbf{z}$ is the target observation based on the noise generation model. $\mathcal{L}_{\text{GAN}}$ is defined as follows:
\begin{equation}
    \begin{aligned}
        \mathcal{L}_{\text{GAN}} &= \mathbb{E}_{\mathbf{M}_{\mathbf{z}}, \mathbf{z}}[\text{log}(D(\mathbf{M}_{\mathbf{z}}, \mathbf{z}))] \\
        & + \mathbb{E}_{\mathbf{M}_{\mathbf{z}}, \mathbf{y}}[\text{log}(1-D(\mathbf{M}_{\mathbf{z}}, G(\mathbf{M}_{\mathbf{z}}, \mathbf{y})))],
        \label{eq:14}
    \end{aligned}
\end{equation}
where $D$ denotes the patch-based discriminator \cite{P2P}, and $G$ denotes the generator, which consists of all the modules in NTGAN for producing $\hat{\mathbf{z}}$.

\section{Experiments}
\subsection{Implementation details}
NTGAN is an unsupervised denoiser, and thus it only requires noisy observations for training. In this work, we mainly consider the realistic noise removal task. Therefore, we evaluate NTGAN on two commonly used real-world denoising benchmarks, i.e. the default 15 cropped noisy images in the Cross-Channel dataset (CC15) \cite{Nam2016AHA} and the Darmstadt Noise dataset (DND) \cite{Plotz2017BenchmarkingDA}. To make a fair comparison with the other denoising methods, we use the cropped images of size $512\times 512$ from SIDD \cite{SIDD}, PolyU \cite{PolyU}, and RENOIR \cite{RENOIR} to form the training set. In the training phase, we randomly crop $8,000$ image patches with size $64 \times 64$ from the noisy images in the training set, and synthesize the training pairs using Eqns. (\ref{eq:9}), (\ref{eq:10}), and (\ref{eq:11}). In addition, the noise variance $\sigma_s$ and $\sigma_c$ in Eqn. (\ref{eq:11}) are uniformly sampled from the range $(0, 0.06]$ and $(0, 0.03]$, respectively. Random rotation and mirroring are applied to the patches for augmentation. It is worth noting that, although the nearly noise-free images are available in SSID , PolyU, and RENOIR, we only use their noisy observations for training. In the testing phase, we feed the noisy observations from DND and CC15 to the trained NTGAN, and set all the elements in $\mathbf{M}_{\mathbf{z}}$ to zero, as we expect NTGAN to generate the noise-free ($\mathbf{M}_{\mathbf{z}} = 0$) images.

We implement NTGAN based on PyTorch \cite{paszke2017automatic}. All the convolutional layers consists of $3\times 3$ kernels with padding $1$ and stride $1$ except for those in the SA units and before the sigmoid function, where the kernel size is $1\times 1$ with padding $0$. The channel number of NTB is fixed to $64$ except for that in the CA units, where $4$ convolutional filters are use. The longest residual connection in the noise branch is linked only for the last 3,000 iterations. We employ Adam \cite{Kingma2014AdamAM} to optimize the objective function defined in Eqn. (\ref{eq:12}) with the trade-off controller $\lambda$ empirically set to $0.3$. We train the network for $10^{6}$ iterations on two Nvidia GEFORCE GTX 1080 Ti GPUs, with batch size $128$, and the learning rate is set to $10^{-4}$ and halved at the $5\times 10^5$th iteration.

\subsection{Evaluation on synthetic AWGN}
To evaluate the denoising performance of NTGAN, we apply it to the synthetic additive white Gaussian noise (AWGN), and compare its performance with the other state-of-the-art denoisers, including BM3D \cite{Dabov2007ImageDB}, WNNM \cite{Gu2014WeightedNN}, DnCNN \cite{Zhang2017BeyondAG}, FFDNet \cite{Zhang2017FFDNetTA}, RIDNet \cite{Anwar_2019_ICCV}, N2N \cite{Lehtinen2018Noise2NoiseLI}, DIP \cite{Ulyanov2017DeepIP}, N2S \cite{N2S}, and N2V \cite{N2V}. To make a fair comparison, we change the original noise generation model, and make it to generate AWGN for producing paired samples in training. Specifically, the training pairs are generated as:
\begin{equation}
    \begin{aligned}
    \mathbf{y} &= \mathbf{x} + \mathbf{n}_{\mathbf{y}}, \text{ \ }\mathbf{z} = \mathbf{x} + \mathbf{n}_{\mathbf{z}}, \\
    \text{ with } \mathbf{n}_{\mathbf{y}} &\sim \mathcal{N}(0, \sigma_{\mathbf{y}}), \text{ \ } \mathbf{n}_{\mathbf{z}} \sim \mathcal{N}(0, \sigma_{\mathbf{z}}),
    \label{eq:4new}
    \end{aligned}
\end{equation}
where both $\sigma_{\mathbf{y}}$ and $\sigma_{\mathbf{z}}$ are uniformly sampled from the range $(0, 75/255]$, and $\mathbf{x}$ represents the clean signal. In other words, the training pairs are obtained by independently adding two small AWGNs to the clean image. We collect all the 4,744 images from the Waterloo Exploration database \cite{ma2017waterloo} to form the training set, and use Eqn. (\ref{eq:4new}) to generate the training pairs. We follow the same settings mentioned in the paper to train up NTGAN and evaluate it on  BSD68 \cite{MartinFTM01} with the noise level set to $15$, $25$, and $50$, respectively. The results are summarized in Table \ref{tab:awgn}. It can be observed that NTGAN achieves comparable or even better results, compared to the supervised non-blind denoisers, i.e. DnCNN and FFDNet. However, NTGAN performs slightly worse than RIDNet when facing AWGN.The reason is that the distribution of AWGN is much easier to be learned, and thus the proposed noise transference strategy loses its advantage by serving as an augmentation approach for learning complicated noise distributions. NTGAN achieves about $31.72$dB, $29.28$dB, and $26.37$dB with the noise level set to 15, 25, and 50, respectively, which basically outperforms the other unsupervised deep denoisers by a large margin.
\begin{table}[ht]
    \centering
    \resizebox{0.98\linewidth}{!}{
    \begin{tabular}{|c|c|c|c|c|c|c|c|c|c|c|}
    \hline
        Type & \multicolumn{2}{c|}{Traditional methods} & \multicolumn{3}{c|}{Supervised CNNs} & \multicolumn{5}{c|}{Unsupervised CNNs} \\
        \hline
        Method & BM3D & WNNM & DnCNN & FFDNet & RIDNet & N2N & DIP & N2S & N2V & NTGAN \\
        \hline
        \multicolumn{11}{|c|}{$\sigma = 15$} \\
        \hline
        PSNR & 31.08 & 31.32 & 31.73 & 31.63 & \textbf{31.81} & \textbf{31.81} & 27.07 & 29.23 & 29.75 & 31.72 \\
        \hline
        \multicolumn{11}{|c|}{$\sigma = 25$} \\
        \hline
        PSNR & 28.57 & 28.83 & 29.23 & 29.23 & \textbf{29.34} & 28.67 & 24.63 & 27.39 & 27.76 & 29.28 \\
        \hline
        \multicolumn{11}{|c|}{$\sigma = 50$} \\
        \hline
        PSNR & 25.62 & 25.83 & 26.23 & 26.29 & \textbf{26.40} & 26.07 & 22.06 & 25.17 & 25.08 & 26.37 \\
        \hline
    \end{tabular}
    }
    \caption{The quantitative results on the grayscale BSD68 images corrupted with AWGN. The best results are highlighted in \textbf{bold}.}
    \label{tab:awgn}
\end{table}

\subsection{Evaluation on realistic noise removal}

\begin{figure*}[t]
    \centering
    \subfigure[Noisy]{
    \begin{minipage}[t]{0.15\linewidth}
    \centering
    \includegraphics[width=\textwidth]{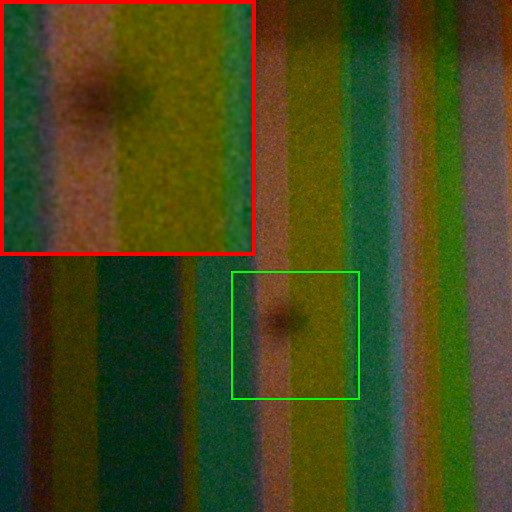}
    \end{minipage}%
    }
    \subfigure[BM3D]{
    \begin{minipage}[t]{0.15\linewidth}
    \centering
    \includegraphics[width=\textwidth]{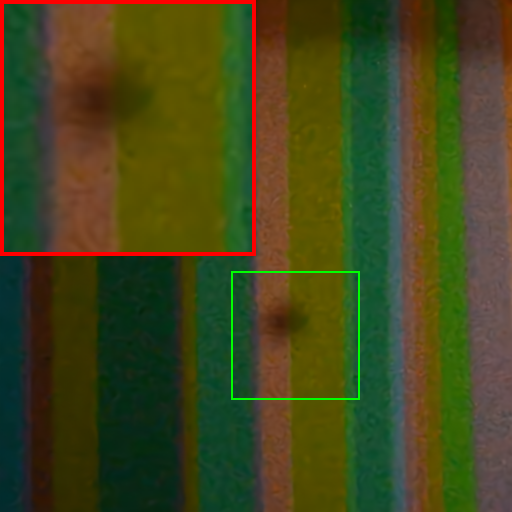}
    \end{minipage}%
    }
    \subfigure[NI]{
    \begin{minipage}[t]{0.15\linewidth}
    \centering
    \includegraphics[width=\textwidth]{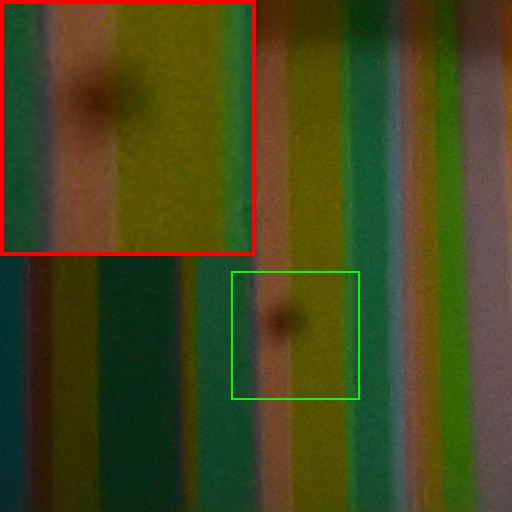}
    \end{minipage}%
    }
    \subfigure[DnCNN+]{
    \begin{minipage}[t]{0.15\linewidth}
    \centering
    \includegraphics[width=\textwidth]{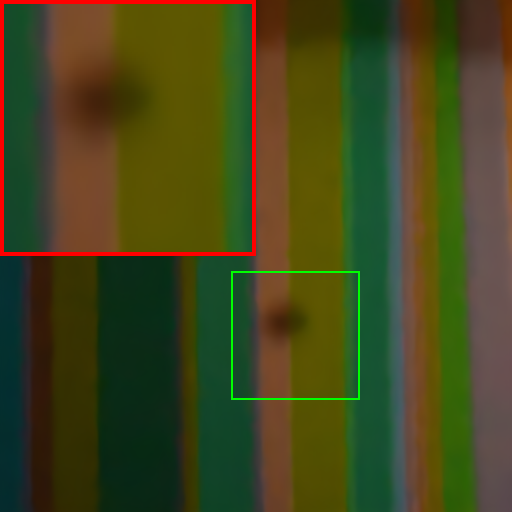}
    \end{minipage}%
    }
    \subfigure[CBDNet]{
    \begin{minipage}[t]{0.15\linewidth}
    \centering
    \includegraphics[width=\textwidth]{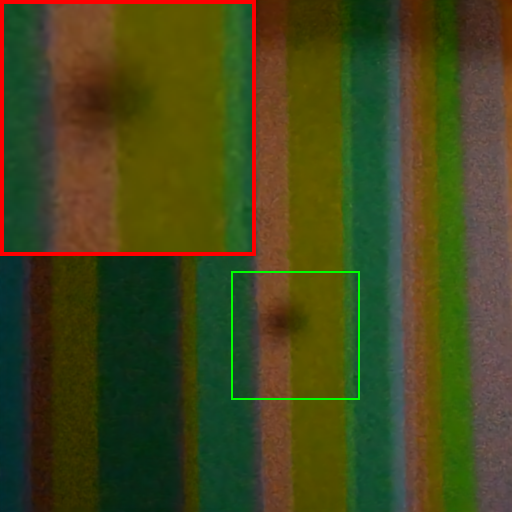}
    \end{minipage}%
    }
    \subfigure[VDNet]{
    \begin{minipage}[t]{0.15\linewidth}
    \centering
    \includegraphics[width=\textwidth]{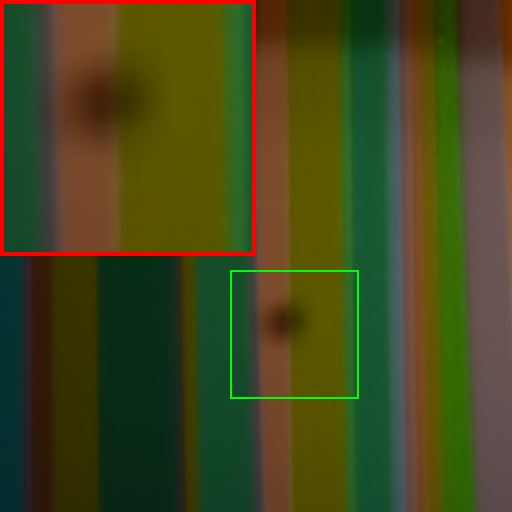}
    \end{minipage}%
    }
    \subfigure[N2N]{
    \begin{minipage}[t]{0.15\linewidth}
    \centering
    \includegraphics[width=\textwidth]{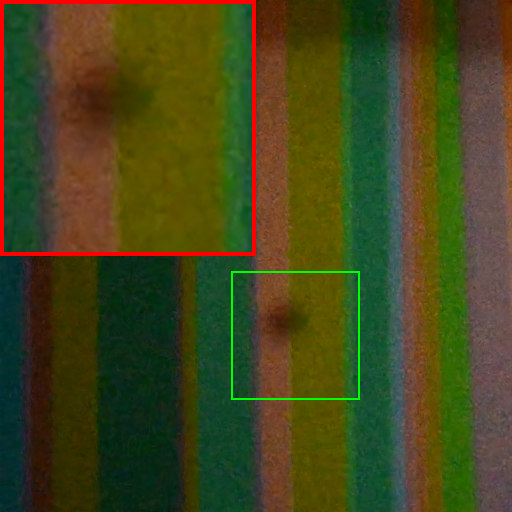}
    \end{minipage}%
    }
    \subfigure[N2S]{
    \begin{minipage}[t]{0.15\linewidth}
    \centering
    \includegraphics[width=\textwidth]{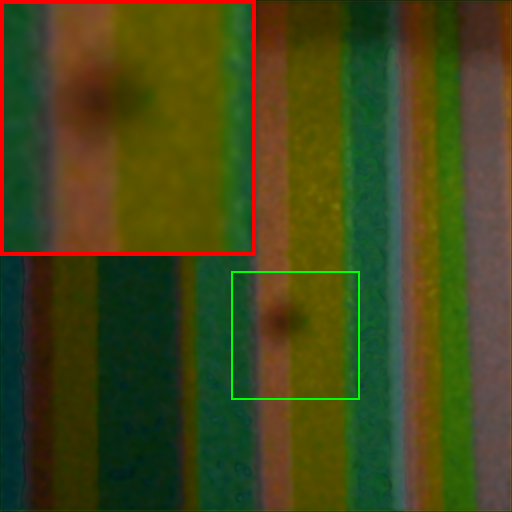}
    \end{minipage}%
    }
    \subfigure[N2V]{
    \begin{minipage}[t]{0.15\linewidth}
    \centering
    \includegraphics[width=\textwidth]{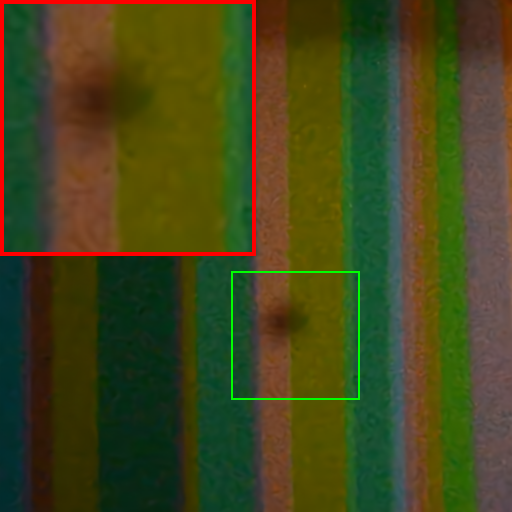}
    \end{minipage}%
    }
    \subfigure[DIP]{
    \begin{minipage}[t]{0.15\linewidth}
    \centering
    \includegraphics[width=\textwidth]{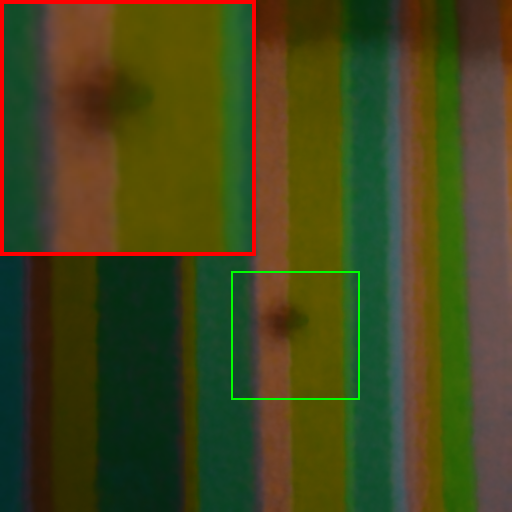}
    \end{minipage}%
    }
    \subfigure[Noise-free]{
    \begin{minipage}[t]{0.15\linewidth}
    \centering
    \includegraphics[width=\textwidth]{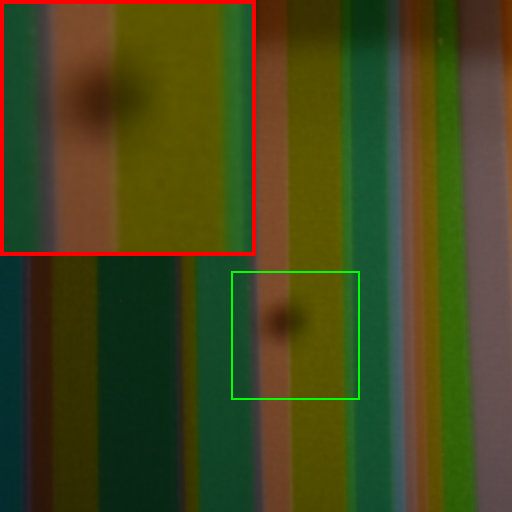}
    \end{minipage}%
    }
    \subfigure[NTGAN]{
    \begin{minipage}[t]{0.15\linewidth}
    \centering
    \includegraphics[width=\textwidth]{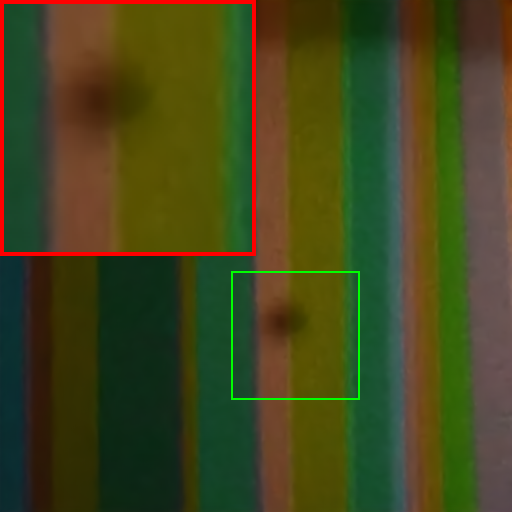}
    \end{minipage}%
    }
    \caption{Evaluation on the perceptual qualities of a real-world noisy image from CC15, restored by different methods.}
    \label{fig:cc15}
\end{figure*}
To evaluate the denoising performance of NTGAN on realistic noise, we applied it to the real noisy photographs in CC15 and DND.

\begin{table}[ht]
    \centering
    \resizebox{0.98\linewidth}{!}{
    \begin{tabular}{|c|c|c|c|c|c|c|c|c|c|c|c|}
    \hline
        Type & \multicolumn{2}{c|}{Traditional method} & \multicolumn{3}{c|}{Supervised} & \multicolumn{6}{c|}{Un/Semi-supervised} \\
        \hline
        Method & BM3D & NI & DnCNN+ & CBDNet & VDNet & N2N & DIP & N2S & N2V & NTGAN & NTGAN$^{*}$ \\
        \hline
        PSNR & 35.19 & 35.33 & 35.40 & 36.44 & 35.84 & 35.32 & 35.69 & 35.38 & 35.27 & 35.74 & \textbf{37.33} \\
        \hline
        SSIM & 0.9063 & 0.9212 & 0.9115 & 0.9460 & 0.9414 & 0.9160 & 0.9259 & 0.9204 & 0.9158 & 0.9243 & \textbf{0.9476}  \\
        \hline
    \end{tabular}
    }
    \caption{The quantitative results on CC15. The best results are highlighted in \textbf{bold}.}
    \label{tab:cc15}
\end{table}

The Cross-Channel dataset (CC15) \cite{Nam2016AHA} consists of eleven noisy images with static
scenes captured by three different cameras. The nearly noise-free ground-truth is obtained by taking the average over the 500 shots with the same camera settings of each observation. We compare our proposed method with the traditional methods, i.e. BM3D \cite{Dabov2007ImageDB} and NI \cite{NeatImage}, the supervised CNN-based methods, i.e. DnCNN+ \cite{Zhang2017BeyondAG}, CBDNet \cite{Guo2018TowardCB}, and VDNet \cite{NIPS2019_8446}, and the state-of-the-art unsupervised methods, i.e. N2N \cite{Lehtinen2018Noise2NoiseLI}, DIP \cite{Ulyanov2017DeepIP}, N2S \cite{N2S} and N2V \cite{N2V}, in terms of the Peak Signal-to-Noise Ratio (PSNR) and the structural similarity (SSIM) index. To make a fair comparison, we adopt the default settings or the pre-trained models provided by the original authors of those compared methods. We also establish the semi-supervised version$^{*}$, which jointly learns from the synthesized NC pairs from \cite{martin2001database} using Eqns. (\ref{eq:9})-(\ref{eq:11}) for learning the augmented noise transference mapping. The quantitative results are listed in Table \ref{tab:cc15}. It can be observed that our proposed method achieves the best performance on CC15. NTGAN$^{*}$ outperforms the supervised networks, which demonstrates that the proposed noise transference strategy can be regarded as a special data augmentation method for the denoising task. NTGAN learns the mapping from one noisy observation to multiple observations with the reliable guidance from $\mathbf{M}_{\mathbf{z}}$, which helps the network to better memorize the noise distributions. Compared with the other unsupervised networks, NTGAN can generally obtain a PSNR gain of about $1$dB, which results from the utilization of the pseudo noise level maps in guiding the learning for different noise distributions. The qualitative comparison is presented in Figure \ref{fig:cc15}. It can be observed that the traditional and the unsupervised competitors tend to retain some noise in the restored image, while DnCNN+ and VDNet oversmooth the image. Although CBDNet generally produces comparable visual results to NTGAN, it creates some distortions on the edges, as shown in the upscaled patches.

\begin{table}[ht]
    \centering
    \resizebox{1\linewidth}{!}{
    \begin{tabular}{|c|c|c|c|c|c|c|c|c|c|c|c|}
    \hline
        Type & \multicolumn{3}{c|}{Traditional method} & \multicolumn{6}{c|}{Supervised} &  \multicolumn{2}{c|}{Un/Semi-supervised} \\
        \hline
        Method & BM3D & WNNM & NI & DnCNN+ & CBDNet & RIDNet & VDNet & GCBD & ADGAN & NTGAN & NTGAN$^{*}$ \\
        \hline
        PSNR & 34.51 & 34.67 & 35.11 & 37.90 & 38.05 & 39.23 & 39.38 & 37.72 & 38.13 & 36.91 & \textbf{39.42} \\
        \hline
        SSIM & 0.8507 & 0.8646 & 0.8778 & 0.9430 & 0.9421 & 0.9526 & 0.9518 & 0.9408 & \textbf{0.9580} & 0.9362 & 0.9560\\
        \hline
    \end{tabular}
    }
    \caption{The quantitative results on DND. The best results are highlighted in \textbf{bold}.}
    \label{tab:dnd}
\end{table}

The Darmstadt Noise dataset (DND) \cite{Plotz2017BenchmarkingDA} is a recently proposed large-scale real-world denoising dataset, which consists of $1,000$ paired noisy and nearly noise-free image patches. The nearly noise-free images are not publicly available, and thus the results can only be obtained through their online submission system. As the above-mentioned unsupervised methods were not evaluated on DND, we just compare NTGAN with the supervised networks and the traditional methods. In addition, GANs have been widely studied and employed for image denoising, as shown in the literature. Thus, we also compare the proposed NTGAN with some state-of-the-art GAN-based denoisers on DND. GAN-based denoisers can be generally categorized into two groups, i.e. using the generator to restore the noisy images, such as ADGAN \cite{Lin_2019_CVPR_Workshops} and using GAN to model the noise generation process for providing training samples, such as GCBD \cite{Chen_2018_cvpr}. The quantitative results are tabulated in Table \ref{tab:dnd}. It can be seen that NTGAN outperforms the other supervised and traditional competitors, which demonstrates its effectiveness in learning the noise transference task as the special augmentation approach for noise removal. Compared to the GAN-based methods, NTGAN gains a PSNR improvement of over $1$dB on DND. This is because the proposed NTGAN utilizes the reliable pseudo supervision from the target noise level in training, which facilitates the learning for different noise distributions. Considering that the clean reference is not required, the proposed strategy has greater potential in practical applications.

\subsection{Evaluation on noise transference}
\begin{figure*}[t]
    \centering
    \subfigure[Original]{
    \begin{minipage}[t]{0.15\linewidth}
    \centering
    \includegraphics[width=\textwidth]{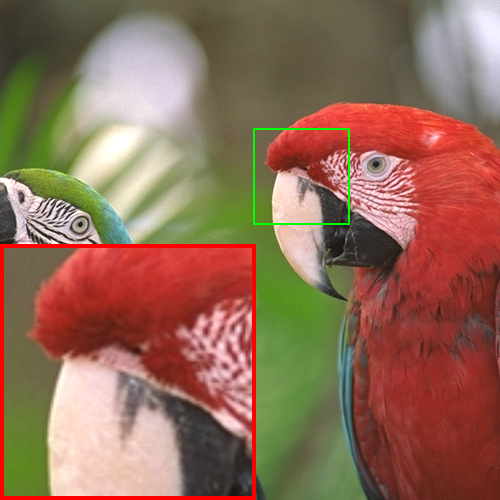} \\
    \vspace{+3pt}
    \includegraphics[width=\textwidth]{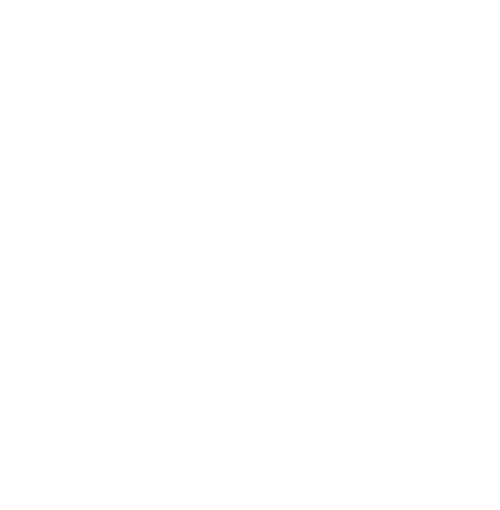}
    \end{minipage}%
    }
    \subfigure[$\mathbf{M}_{\mathbf{z}}=15$]{
    \begin{minipage}[t]{0.15\linewidth}
    \centering
    \includegraphics[width=\textwidth]{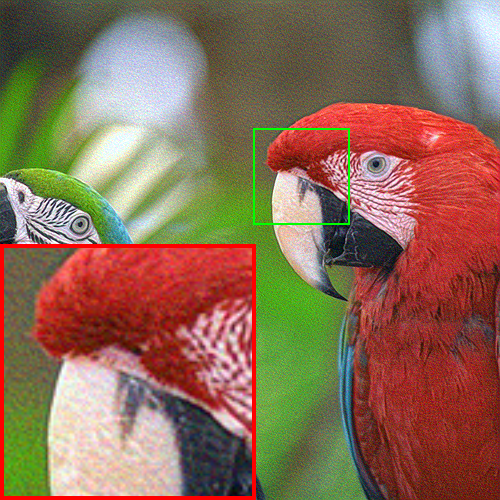} \\
    \vspace{+3pt}
    \includegraphics[width=\textwidth]{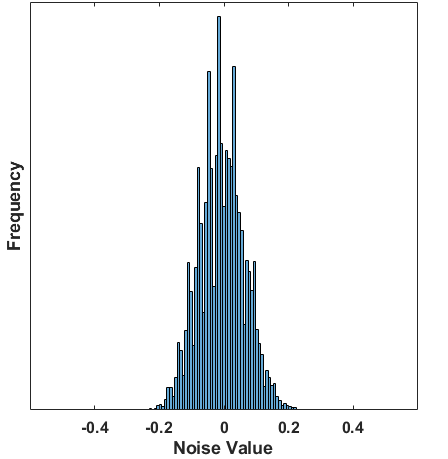}
    \end{minipage}%
    }
    \subfigure[$\mathbf{M}_{\mathbf{z}}=25$]{
    \begin{minipage}[t]{0.15\linewidth}
    \centering
    \includegraphics[width=\textwidth]{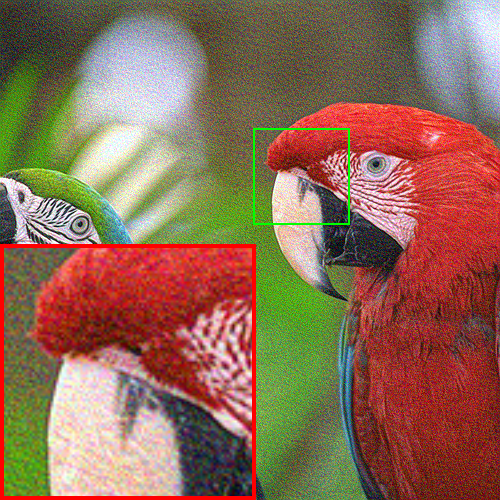} \\
    \vspace{+3pt}
    \includegraphics[width=\textwidth]{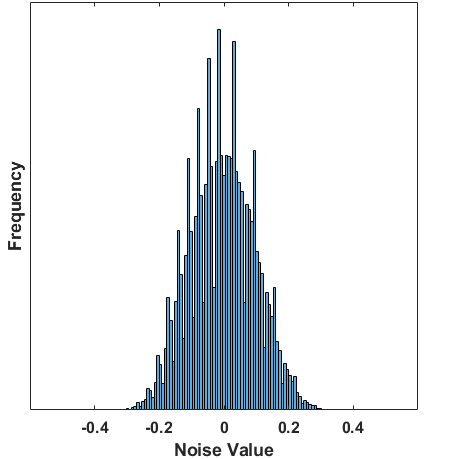}
    \end{minipage}%
    }
    \subfigure[$\mathbf{M}_{\mathbf{z}}=50$]{
    \begin{minipage}[t]{0.15\linewidth}
    \centering
    \includegraphics[width=\textwidth]{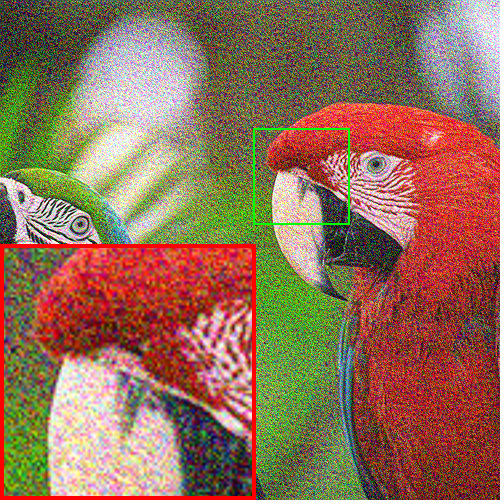} \\
    \vspace{+3pt}
    \includegraphics[width=\textwidth]{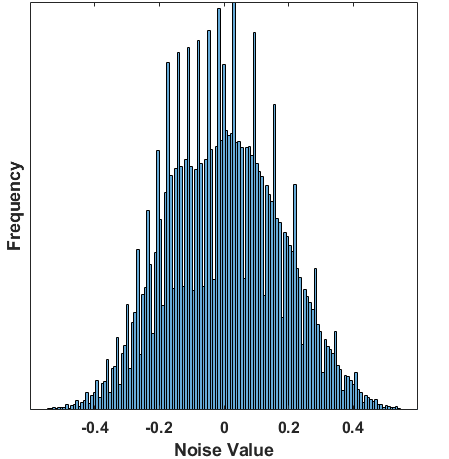}
    \end{minipage}%
    }
    \subfigure[Guidance]{
    \begin{minipage}[t]{0.15\linewidth}
    \centering
    \includegraphics[width=\textwidth]{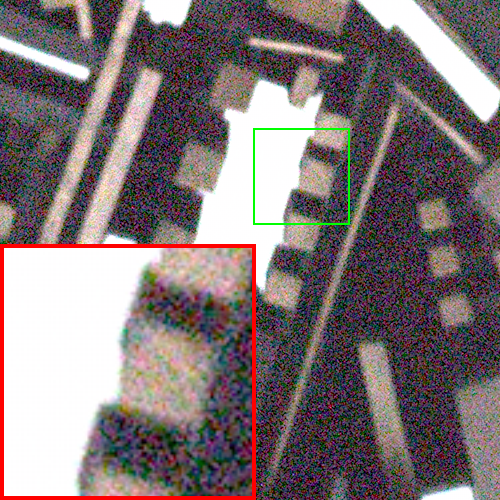} \\
    \vspace{+3pt}
    \includegraphics[width=\textwidth]{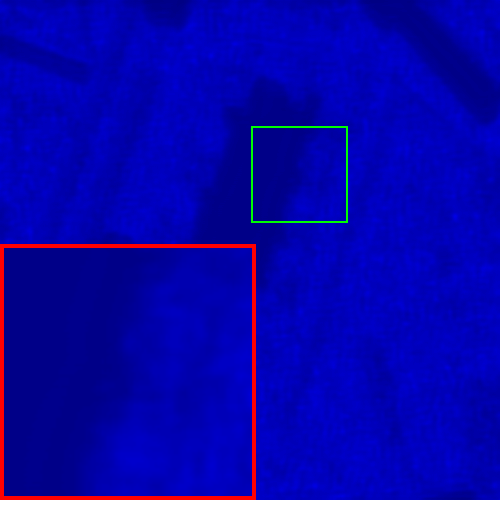}
    \end{minipage}%
    }
    \subfigure[Results]{
    \begin{minipage}[t]{0.15\linewidth}
    \centering
    \includegraphics[width=\textwidth]{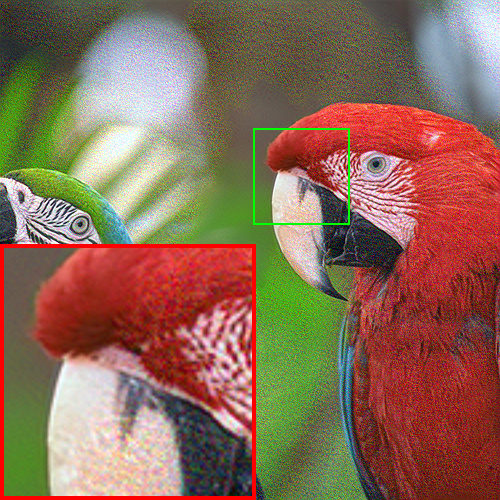} \\
    \vspace{+3pt}
    \includegraphics[width=\textwidth]{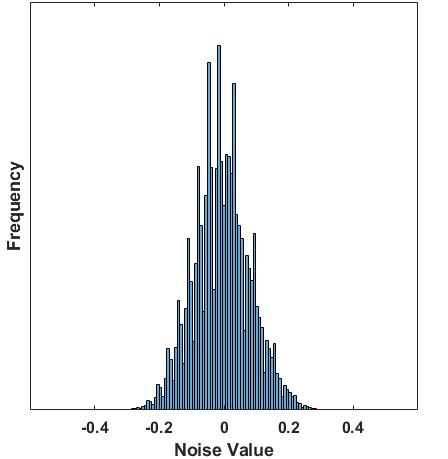}
    \end{minipage}%
    }
    \caption{The evaluation on noise transference. In the first row, (b)-(d) are the synthetic noisy images based on the uniform noise level maps, (e) presents the guiding image selected from DND, and (f) presents the transference result based on the noise level map extracted from (e). In the second row, (b)-(d) and (f) present the noise distribution in the synthetic images, and (e) presents the estimated noise level map of the guiding image.}
    \label{fig:nt}
\end{figure*}
To evaluate the performance of NTGAN for noise transference, we consider two cases, i.e. transferring the synthetic AWGN and transferring the realistic noise. As NTGAN synthesizes the target observation based on the guiding noise level map, we employ the pre-trained noise level estimation sub-network in CBDNet \cite{Guo2018TowardCB} to extract the target $\mathbf{M}_{\mathbf{z}}$ from a DND image for guiding the realistic noise transference task. The results are presented in Fig. \ref{fig:nt}. For simplicity, we set the input observation as a clean image. It can be observed from Figs. \ref{fig:nt}(b)-\ref{fig:nt}(d) that when $\mathbf{M}_{\mathbf{z}}$ increases, NTGAN introduces stronger noise into the synthetic image. More importantly, when guided by the realistic noise level map in Fig. \ref{fig:nt}(e), the noise in the transferred image follows the same pattern as that in the guiding image, which demonstrates the effectiveness of NTGAN in memorizing the signal-dependent noise.

\subsection{Ablation study}
\begin{table}[ht]
    \centering
    \resizebox{0.95\linewidth}{!}{
    \begin{tabular}{|c|c| c |c |c |c |c |c| c|c|}
        \hline
        $\mathcal{L}_{\text{GAN}}$ & \checkmark & \checkmark & \checkmark & \checkmark & \checkmark & &\checkmark &\checkmark &\checkmark\\
        RoR &  & \checkmark & \checkmark & \checkmark &  & \checkmark &\checkmark &\checkmark &\checkmark\\
        CA &  &  &\checkmark &  & \checkmark & \checkmark & \checkmark &\checkmark &\checkmark\\
        SA &  &  &  & \checkmark & \checkmark & \checkmark & \checkmark &\checkmark &\checkmark\\
        N2C &  &  &  & &  &  & &\checkmark &\\
        N2N &  &  &  & &  &  & & & \checkmark\\
        \hline
        \hline
        CC15 & 35.50 & 36.41 & 36.73 & 36.67 & 35.88 & 37.25 & 37.33 & 37.21 & 36.52\\
        \hline
        DND & 37.68 & 38.06 & 38.40 & 38.58 & 37.87 & 39.33 & 39.42 & 39.36 & 38.49\\
        \hline
    \end{tabular}}
    \caption{The ablation study on DND and CC15 in term of PSNR. $\mathcal{L}_{\text{GAN}}$: GAN loss. RoR: residual-on-residual structure. CA: channel attention unit. SA: spatial attention unit. N2C: noise-to-clean manner. N2N: noise-to-noise manner.}
    \label{tab:as}
\end{table}
To provide a comprehensive analysis, we perform ablation studies to investigate the effectiveness of the different designs. Specifically, we consider the effect of the GAN loss, the residual-on-residual structure (RoR), the channel attention unit, and the spatial attention unit in NTGAN. In addition, we also consider training the network in a noise-to-clean (N2C) manner and a noise-to-noise (N2N) manner. Specifically, N2C uses the clean references as the target images for training, while N2N detaches the noise-level-encoding branch and fits the noisy observations without the guidance from the target noise level maps. To make a fair comparison, we enlarge the kernel size and the network depth of the models without CA, SA, or the noise-level-encoding branch, in order to make their model capacity equal to, or larger than, the original NTGAN. It can be observed from Table \ref{tab:as} that the denoising performance is slightly degraded when omitting $\mathcal{L}_{\text{GAN}}$. Moreover, we observe that the network, trained without $\mathcal{L}_{\text{GAN}}$, fails to perform accurate noise transference. If we increase the value of the target noise level, the network produces more blurred images. Both attention units contribute to a PSNR improvement of about 1dB and 0.5dB on DND and CC15, respectively. The residual-on-residual (RoR) structure plays an important role in blind denoising. Without RoR, the performance is dramatically degraded, because the network cannot flexibly deal with the input observations with different noise levels. When N2C is applied, the results are slightly degraded, compared to the noise transference strategy. This is because noise transference, as introduced in the previous section, can serve as a special augmentation approach for the denoising task. Moreover, when the model is trained in an N2N manner, the performance is dramatically degraded. The proposed noise transference strategy utilizes the guidance from the noise intensity at the different positions of the observation in training, which facilitates the learning for the distribution of the spatially variant noise, and consequently enhances the model generalization on real photographs. Therefore, the proposed noise transference strategy is beneficial to the denoising performance.

\section{Conclusion}
In this paper, we have proposed a novel noise transference generative adversarial network, namely NTGAN, for blindly removing the realistic noise residing in photographs. Our proposed method treats the noise transference task as the general form of the noise reduction task, which enables it to obtain the denoising ability by observing the corrupted data only. NTGAN learns the noise transference task based on the proposed pseudo noise level strategy, which provides the reliable guidance for learning different noise distributions. We have evaluated NTGAN on two widely used real-world denoising benchmarks. The experimental results have showed that our proposed method achieves promising performance on real photographs, making it a potential solution to practical denoising problems.

\bibliographystyle{IEEEtran.bst}
\bibliography{bibfile}

\end{document}